\documentclass[10pt]{article}
\usepackage[margin=2cm]{geometry}
\usepackage{cite}
\usepackage{geometry}                % See geometry.pdf to learn the layout options. There are lots.
\usepackage{graphicx}
\usepackage{amssymb}
\usepackage{verbatim}
\usepackage{amsmath,amssymb}
\usepackage{epsfig}
\usepackage{graphicx}
\usepackage[usenames,dvipsnames]{color}

\title{
\vspace*{-2cm}
\begin{flushright}
\normalsize{EFI-13-7 \\
ANL-HEP-PR-13-25}
~\\
\end{flushright}
\vspace*{1.5cm}
Electroweak Baryogenesis with a Supersymmetric Sector\\
\author{\textbf{Ran Huo} \\
~\\
\normalsize\emph{Enrico Fermi Institute, University of Chicago, Chicago, IL 60637}
}}
\begin{document}
\maketitle
\vspace*{0.5cm}
\begin{abstract}
We study a model with an exotic new sector strongly coupled to the Higgs boson, in which supersymmetry is introduced to protect the quartic coupling from a large running and avoid potential vacuum stability problem. The fermionic components present vector like mass terms, through which the Higgs diphoton decay branching ratio can be tuned. The bosonic components trigger a strongly first order electroweak phase transition. We find a large parameter region of effective Yukawa coupling $y\gtrsim2$ and mass parameters $m_f\sim m_s$ of a few hundred GeV, that can simultaneously accommodate the diphoton excess and electroweak baryogenesis, without vacuum stability and electroweak precision measurement problems.
\end{abstract}
\thispagestyle{empty}

\noindent\textbf{Introduction.} The Large Hadron Collider (LHC) experiments have reported the discovery of a new particle, with properties similar to the ones of the Standard Model (SM) Higgs boson, and with a mass of roughly $125$~GeV \cite{ATLAS:2012ae, Chatrchyan:2012tx, Aad:2012tfa, Chatrchyan:2012ufa}. However, the early day best fit for the rate of the diphoton production channel, proceeding from Higgs production and decay, seems to allow departure from the SM prediction. While the ATLAS experiment still sees an excess \cite{ATLAS:2013oma}
\begin{equation}\label{ATLAS}
\frac{\left[\sigma(pp\rightarrow h)\times BR(h\rightarrow\gamma\gamma)\right]_{\text{best-fit}}}{\sigma_{\text{SM}}(pp\rightarrow h)\times BR_{\text{SM}}(h\rightarrow\gamma\gamma)}=1.65\pm0.24^{+0.25}_{-0.18},
\end{equation}
with the recent updates CMS doesn't see the previous excess \cite{ATLAS:2012ae, Chatrchyan:2012tx, Aad:2012tfa, Chatrchyan:2012ufa} anymore \cite{CMS:ril}
\begin{equation}\label{CMS}
\frac{\left[\sigma(pp\rightarrow h)\times BR(h\rightarrow\gamma\gamma)\right]_{\text{best-fit}}}{\sigma_{\text{SM}}(pp\rightarrow h)\times BR_{\text{SM}}(h\rightarrow\gamma\gamma)}=0.78 ^{+0.28}_{-0.26} \;\;  (1.11 ^{+0.32}_{-0.30}),
\end{equation}
A naive combination of the two experiments results lead to no statistically significant deviation from the SM. But it is still interesting to study the possibility of a deviation from the SM prediction. On the other hand, the preliminary indications are that the top induced gluon fusion Higgs production rate and the $h\rightarrow Z Z^*$ and $h\rightarrow W W^*$ branching ratio are in better alignment with the SM expectations. Given that the Higgs diphoton amplitude is dominated by $W^\pm$ and top loops in the SM, the only possibility is that new light charged particle interferes with the $W^\pm$ and top loops.

Such an exotic particle must couple to the Higgs boson strongly, and it is exactly a new candidate helping to trigger electroweak (EW) baryogenesis \cite{Quiros:1999jp, Trodden:1998ym}. Successful EW baryogenesis could be obtained in the presence of a strongly first order phase transition, in which the Higgs vacuum expectation value (VEV) jumps from the symmetric value $\langle\phi\rangle=0$ to some nonzero value (not necessarily the zero temperature value $v=246.2$~GeV), through a quantum tunneling effect. First order phase transition means a potential barrier between the high temperature symmetric phase and the $\langle\phi\rangle\neq0$ local minimum. The quantum tunneling phase transition is described by a bubble nucleation process in real space; while ``strongly'' means \cite{Shaposhnikov}
\begin{equation}\label{strong1st}
\frac{\langle\phi(T_n)\rangle}{T_n}\gtrsim1,
\end{equation}
where $T_n$ is the temperature for nucleation, or the phase transition. This jump of the Higgs VEV induces a jump of the weak gauge boson mass through the EW symmetry breaking mechanism, and the previous copious sphaleron process will suddenly be terminated by a Boltzmann suppression $e^{-\frac{E_s}{T}}=e^{-\frac{8\pi m_W(\phi)}{g^2T}B(\frac{\lambda}{g^2})}$. If some CP violation source exists and results in some baryonic asymmetry during the nucleation, Eq.~(\ref{strong1st}) will prevent the generated asymmetry from being washed out, so it is the necessary condition for EW baryogenesis and the goal for model building.

It is well known that the SM with a $125$~GeV Higgs boson is insufficient for a strongly first order phase transition \cite{Carrington:1991hz, Dine:1992vs, Arnold:1992fb, Espinosa:1992kf, Buchmuller:1993bq}. In finite temperature quantum field theory, the Higgs effective potential is corrected not only by the zero temperature loop effects, but also by the thermal loops. In the high temperature expansion of thermal loops, bosons directly contribute to cubic terms, which strengthen the EW phase transition. On the other hand, fermions do not directly contribute such terms, but some less efficient effect may exist such as suggested in \cite{Carena:2004ha, Davoudiasl:2012tu}. In general only strongly coupled new particles are important to EW baryogenesis, such as top squark \cite{Carena:1996wj}. However, the light stop scenario itself is severely constrained by the LHC experiments \cite{Curtin:2012aa, Carena:2012np}.

In this paper we design a minimal model which simultaneously accommodates the diphoton excess and the EW baryogenesis, aiming to figure out the parameter region. Both the diphoton excess and EW baryogenesis require a large coupling to Higgs, and it may induce large corrections in the quartic coupling renormalization group equation (RGE) running. If the new particle is a fermion, the quartic coupling RGE beta coefficient is negative, and the quartic coupling may eventually run negative at certain high scale, rendering the vacuum unbounded from below. To avoid this problem the new sector is supersymmetric, in that there are equal number of complex scalars and chiral fermions with the same coupling\footnote{Regardless of the nature we will call it Yukawa coupling.} $y$ to the Higgs boson, so that their contribution to the quartic coupling beta function get canceled above threshold. %This can be embedded into a chiral supermultiplet, but their mass may have other contributions in addition to the part proportional to Higgs VEV.

\noindent\textbf{The model.} In more detail our model is the supersymmetric extension of \cite{ArkaniHamed:2012kq}. The anchor is, the \emph{charged} fermionic part has $n_\pm$ complex degree of freedom with mass matrix
\begin{equation}
{\cal M}_{F^\pm}=\left(\begin{array}{cc}
m_\psi & \frac{1}{\sqrt{2}}y\phi \\
\frac{1}{\sqrt{2}}y\phi & m_\chi
\end{array}\right),\label{chargedvectorfermion}
\end{equation}
where $m_\psi,m_\chi$ are vector like mass terms. The eigenvalues are $M_{F_{1,2}^\pm}=\frac{1}{2}\Big(m_\psi+m_\chi\mp\sqrt{(m_\psi-m_\chi)^2+2y^2\phi^2}\Big)$. The mass eigenstate $(F_1^\pm, F_2^\pm)^T$ %(or the original $(\psi^\pm, \chi^\pm)^T$)
can be obtained by a rotation
\begin{equation}\label{MassInteraction}
\left(\begin{array}{c}
F_1^\pm \\
F_2^\pm
\end{array}\right)=\left(\begin{array}{cc}
\cos\theta & -\sin\theta \\
\sin\theta & \cos\theta
\end{array}\right)\left(\begin{array}{c}
\psi^\pm \\
\chi^\pm
\end{array}\right),
\end{equation}
where the mixing angle satisfies
\begin{equation}\label{mixingangle}
\tan2\theta=\frac{\sqrt{2}yv}{m_\chi-m_\psi}.
\end{equation}
We have two Dirac mass eigenstates $F_1^\pm$ and $F_2^\pm$ with $n_\pm=4$ for each of them, so the total number of degree of freedom $n_\pm$ is 8 for a minimum model.

Embedding it into the $SU(2)_L\times U(1)_Y$ electroweak representation leads to two kinds of models, namely the mirror leptonic model \cite{Joglekar:2012vc, An:2012vp, Kearney:2012zi, Davoudiasl:2012ig, Bae:2012ir, Voloshin:2012tv, McKeen:2012av, Lee:2012wz, Arina:2012aj, Batell:2012zw, Fan:2013qn, Carmona:2013cq, Cheung:2013bn, Feng:2013mea, Englert:2013tya} and the wino-higgsino model \cite{Huo:2012tw}. In the leptonic case $\psi_{L,R}\sim(1,2)_{-\frac{1}{2}}$ and $\chi_{L,R}\sim(1,1)_{-1}$\footnote{In addition to the charged lepton we should also have neutrinos \cite{Witten:1982fp}, namely additional term $-y_n\psi^c_Li\sigma_2\phi\chi'_R-y_n\chi'^c_L \phi^ci\sigma_2\psi_R$ with $\chi'\sim(1,1)_{0}$ in the Lagrangian. This neutral particle doesn't contribute to the diphoton amplitude. If $y_n$ is small then its contribution to the Higgs quartic coupling running is negligible, and so are the contribution to electroweak precision measurement observables. Therefore we ignore them in the calculation.}. The electrical neutral fermion is only $\phi^0$ with a mass $m_\psi$, it doesn't couple to Higgs so $M_{F^0}=m_\psi$ and $n_0=0$ where $n_0$ is the neutral degree of freedom coupling to the Higgs boson. In order to cancel the chiral anomaly we need the new sector to be mirror-like or have both left handed and right handed chiral fermions, which correctly reproduces the two charged Dirac fermions. The relevant Lagrangian is
\begin{equation}
\mathcal{L}\supset-y\psi^c_L \phi\chi_R-y\chi^c_L \phi^c\psi_R-m_\psi\psi^c_L\psi_R-m_\chi\chi^c_L\chi_R+h.c..
\end{equation}
On the other hand, in the wino-higgisno model $\psi\sim(1,2)_{\pm\frac{1}{2}}$ and $\chi\sim(1,3)_{0}$. Just like in the MSSM case we need two higgsinos $\psi$s to reproduce $F_1^\pm$ and $F_2^\pm$. They could be exactly the MSSM higgsino case with hypercharges opposite to each other, the Lagrangian of which is
\begin{equation}
\mathcal{L}\supset-\sqrt{2}y\psi^c_1\chi i\sigma_2\phi-\sqrt{2}y\phi^ci\sigma_2\chi\psi_1-\sqrt{2}y\psi^c_2\chi \phi-\sqrt{2}y\phi^c\chi\psi-m_\psi(\psi_1i\sigma_2\psi_2+\psi_2i\sigma_2\psi_1)-m_\chi\chi\chi.
\end{equation}
And the induced neutral fermion mass mixing matrix is similar to the MSSM neutralino one
\begin{equation}\label{neutralvectorfermion1}
{\cal M}_{F^0}=\left(\begin{array}{ccc}
0 & -m_\psi & \frac{1}{2}y\phi \\
-m_\psi & 0 & -\frac{1}{2}y\phi \\
\frac{1}{2}y\phi & -\frac{1}{2}y\phi & m_\chi
\end{array}\right),
\end{equation}
with the mass eigenstates $M_{F_1^0}=m_\psi$, $M_{F_{2,3}^0}=\frac{1}{2}\Big(m_\psi+m_\chi\mp\sqrt{(m_\psi-m_\chi)^2+2y^2\phi^2}\Big)=M_{F_{1,2}^\pm}$. Note that $F_1^0$ doesn't couple to Higgs, so the total neutral degree of freedom coupling to the Higgs boson $n_0=4$ comes from two Majorana fermion $F_{2,3}^0$. This setup is consistent in the supersymmetric case with a two higgs doublet model of different Higgs hypercharges $\pm\frac{1}{2}$.

On the other hand, we simply assume there are $n_\pm$ charged and $n_0$ neutral degrees of freedom for gauge bosons or scalars with mass square $M_S^2=m_s^2+\frac{1}{2}y^2\phi^2$, where $m_s$ is the scalar soft mass. We only add the new naively supersymmetric sector to the SM, that completes our model.
\bigskip

\noindent\textbf{The diphoton amplitude.} The low energy theorem \cite{Ellis:1975ap} gives a good estimation of the Higgs diphoton loop amplitude \cite{Carena:2012xa}
\begin{equation}\label{LowETheorem}
{\cal L}_{h\gamma\gamma}\simeq\frac{\alpha}{16\pi}\frac{h}{v}\frac{\partial}{\partial\ln v}\sum_ib_iQ_i^2\ln\left(\det{\cal M}_i^\dag{\cal M}_i\right)F_{\mu\nu}F^{\mu\nu},
\end{equation}
where $\alpha\equiv\frac{e^2}{4\pi}$ is the fine structure constant, $b_i$ is the contribution for particle $i$ to the QED beta function through the photon vacuum polarization diagram, $Q_i$ is the electric charge in unites of $e$, and $\mathcal{M}_i$ is the mass matrix. In the SM the Higgs diphoton decay amplitude is dominated by $W^\pm$ and top loops, because only the two have significant coupling to the Higgs boson through the $SU(2)$ gauge coupling $g\simeq0.65$ and top Yukawa coupling $y_t=\sqrt{2}m_{t,run}/v\simeq0.95$. Among the two the $W^\pm$ loop is dominant, giving a contribution proportional to $-8.32$ for a $125$~GeV SM Higgs boson, while the top loop contribution is $1.84$ leading to a destructive interference\footnote{The above low energy theorem approximation gives separately $-\frac{22}{3}=-7.33$ and $\frac{16}{9}=1.78$}. Note that a \emph{chiral} fermion contributes to the one loop QED beta function by $b_F=\frac{2}{3}$ and a \emph{complex} scalar contributes to the one loop QED beta function by $b_S=\frac{1}{3}$. Normalizing the new amplitude contribution we have \cite{ArkaniHamed:2012kq}
\begin{equation}\label{diphoton}
\mu_{\gamma\gamma}=\bigg|1-\frac{1}{6.48}\frac{2}{3}\frac{\partial\ln(\det{\cal M}_{F^\pm}^\dag{\cal M}_{F^\pm})}{\partial\ln v}-\frac{1}{6.48}\frac{1}{3}\frac{\partial\ln(\det{\cal M}_{S^\pm}^2)}{\partial\ln v}\bigg|^2.
\end{equation}
where $\mu_{\gamma\gamma}$ is the diphoton channel signal strength or the branching ratio enhancement factor. We have taken electrical charge $Q=1$. Note that corresponding to the $2\times2$ fermion mass matrix ${\cal M}_{F^\pm}$, here the boson mass matrix should also be $2\times2$ (originally in the basis of $\bar{\psi}$ and $\bar{\chi}$).

%Note that in the fermion part we introduce an additional factor of $\frac{1}{2}$, that's because the fermion has a 2 by 2 mass matrix Eq.~(\ref{chargedvectorfermion}), and through determinant and logrithm it will introduce a factor of 2, the $\frac{1}{2}$ is introduced to cancel it. The factor $2$ in the bosonic part is because one Dirac fermion corresponds to two complex scalars by supersymmetry.

The bosonic contribution is $\partial\ln(\det{\cal M}_S^2)/\partial\ln v=2y^2v^2/(2m_s^2+y^2v^2)$, positive but always smaller than two, so its effect is always to decrease the diphoton rate. On the other hand, the fermionic contribution is $\partial\ln(\det{\cal M}_S^2)/\partial\ln v=-2y^2v^2/(2m_\psi m_\chi-y^2v^2)<0$ if $m_\psi m_\chi>\frac{1}{2}y^2v^2$, which gives the probably desired diphoton enhancement. This is the idea in many diphoton excess model buildings \cite{ArkaniHamed:2012kq, Joglekar:2012vc, Huo:2012tw, An:2012vp, Kearney:2012zi, Davoudiasl:2012ig, Bae:2012ir, Voloshin:2012tv, McKeen:2012av, Lee:2012wz, Arina:2012aj, Batell:2012zw, Fan:2013qn, Carmona:2013cq, Cheung:2013bn, Feng:2013mea, Englert:2013tya}, the vector like mass term flips the sign dependence of the product of the two mass eigenstates in derivative with the Higgs VEV. Here we parameterize the model by setting $\mu_{\gamma\gamma}=(1+\Delta{\cal A}_F-\Delta{\cal A}_B)^2$ where $\Delta{\cal A}_F$ and $\Delta{\cal A}_B$ are separately the fractional fermionic and bosonic diphoton amplitude change, then we choose to solve the mass parameters from $\Delta{\cal A}_F$
\begin{equation}\label{Mf0}
m_\psi m_\chi=\frac{1}{2}y^2v^2\left(1+\frac{n_\pm}{19.44\Delta{\cal A}_F}\right),
\end{equation}
and with $\cot2\theta$ we get
\begin{equation}\label{Msoftf}
m_{\psi,\chi}=\Big(\sqrt{1+\cot^22\theta+\textstyle{\frac{n_\pm}{19.44\Delta{\cal A}_F}}}\mp\cot2\theta\Big)\frac{yv}{\sqrt{2}}.
\end{equation}
The fermion mass eigenstates with arbitrary Higgs VEV $\phi$ are
\begin{eqnarray}\label{Mf}
M_{F_{1,2}^\pm}=M_{F_{2,3}^0}\hspace{-0.7em}&=&\hspace{-0.7em}\Big(\sqrt{1+\cot^22\theta+\textstyle{\frac{n_\pm}{19.44\Delta{\cal A}_F}}}\mp\sqrt{\textstyle{\frac{\phi^2}{v^2}}+\cot^22\theta}\Big)\frac{yv}{\sqrt{2}},\nonumber\\
M_{F^0}=m_\psi\hspace{-0.7em}&=&\hspace{-0.7em}\Big(\sqrt{1+\cot^22\theta+\textstyle{\frac{n_\pm}{19.44\Delta{\cal A}_F}}}-\cot2\theta\Big)\frac{yv}{\sqrt{2}}.
\end{eqnarray}
\bigskip

\noindent\textbf{The zero temperature potential.} The tree level Higgs potential is
\begin{equation}\label{V0}
V_0=-\frac{1}{4}m_h^2\phi^2+\frac{1}{4}\lambda\phi^4,
\end{equation}
where $m_h=125$~GeV and $\lambda=\frac{m_h^2}{2v^2}=0.1291$.

The one loop corrections are also dominated by particles which couple to the Higgs boson strongly, namely the $W^\pm$, $Z^0$ gauge bosons and the top quarks in the SM, as well as our new bosons and fermions. The zero temperature one loop corrections are given by
\begin{equation}\label{generalV1}
V_{1,i}=\pm\frac{n_i}{64\pi^2}\big(m_i^2(\phi)\big)^2\left(\ln\frac{m_i^2(\phi)}{\Lambda^2}-c\right),
\end{equation}
where ``$+$'' is for boson and ``$-$'' is for fermion, $\Lambda$ is the renormalization scale, and $c=\frac{5}{6},\frac{3}{2}$ separately for gauge boson and all the other particles.

The above expression is unrenormalized. By requiring the tree level minimum and Higgs mass not to be shifted by one loop correction we can impose renormalization condition $\frac{dV_1}{d\phi}\big|_{\phi=v}=0$ and $\frac{d^2V_1}{d\phi^2}\big|_{\phi=v}=0$, and we get a renormalized one loop correction
\begin{eqnarray}\label{renormalizedV1}
\bar{V}_{1,i}&=&\pm\frac{n_i}{64\pi^2}\Bigg(\big(m^2_i(\phi)\big)^2\ln m^2_i(\phi)+\bigg[\Big(-\frac{3}{4}\frac{m^2_i(v)m^2_i(v)'}{v}+\frac{3}{4}\big(m^2_i(v)'\big)^2+\frac{1}{4}m^2_i(v)m^2_i(v)''\Big)\nonumber\\
&&+\Big(-\frac{3}{2}\frac{m^2_i(v)m^2_i(v)'}{v}+\frac{1}{2}\big(m^2_i(v)'\big)^2+\frac{1}{2}m^2_i(v)m^2_i(v)''\Big)\ln m^2_i(v)\bigg]\phi^2\nonumber\\
&&+\bigg[\Big(\frac{1}{8}\frac{m^2_i(v)m^2_i(v)'}{v}-\frac{3}{8}\big(m^2_i(v)'\big)^2-\frac{1}{8}m^2_i(v)m^2_i(v)''\Big)\nonumber\\
&&+\Big(\frac{1}{4}\frac{m^2_i(v)m^2_i(v)'}{v}-\frac{1}{4}\big(m^2_i(v)'\big)^2-\frac{1}{4}m^2_i(v)m^2_i(v)''\Big)\ln m^2_i(v)\bigg]\frac{\phi^4}{v^2}\Bigg)\\
&\to&\pm\frac{n_i}{64\pi^2}\bigg(\big(m_i^2(\phi)\big)^2\left(\ln\frac{m_i^2(\phi)}{m_i^2(v)}-\frac{3}{2}\right)+2m_i^2(v)m_i^2(\phi)\bigg).
\end{eqnarray}
The first equation is for a generic mass square as a function of $\phi^2$ so that the counterterm in the effective potential is up to $\phi^4$, while the second one is assuming $m^2_i(\phi)=a+b\phi^2$. All the SM particles as well as the new boson will follow the second, but our new fermion can only be described by the first one. We have summarized all the relevant quantities in Table \ref{V1particlelist}.

\begin{table}
\centering
\begin{tabular}{ccccc}
Particle & $\pm$ & $n_i$ & $m_i(\phi)^2$ & $\Pi_T$ \\
\hline
$h$ & $+$ & $1$ & $\lambda(3\phi^2-v^2)$ & $\Big(\frac{1}{2}\lambda+\frac{1}{4}y_t^2+\frac{1}{16}g'^2+\frac{3}{16}g^2\Big)T^2$ \\
$G$ & $+$ & $3$ & $\lambda(\phi^2-v^2)$ & $\Big(\frac{1}{2}\lambda+\frac{1}{4}y_t^2+\frac{1}{16}g'^2+\frac{3}{16}g^2\Big)T^2$ \\
$W_T^\pm$ & $+$ & $4$ & $\frac{1}{4}g^2\phi^2$ & $0$ \\
$W_L^\pm$ & $+$ & $2$ & $\frac{1}{4}g^2\phi^2$ & $\frac{11}{6}g^2T^2$ \\
$Z_T^0$ & $+$ & $2$ & $\frac{1}{4}(g^2+g'^2)\phi^2$ & $0$ \\
$Z_L^0$ & $+$ & $1$ & $\frac{1}{4}(g^2+g'^2)\phi^2$ & $\frac{11}{6}g'^2T^2$ \\
$t$ & $-$ & $12$ & $\frac{1}{2}y_t^2\phi^2$ & $0$ \\
\hline
$F_1^\pm$ & $-$ & $4$ & In Eq.~\ref{Mf} & $0$ \\
$F_2^\pm$ & $-$ & $4$ & In Eq.~\ref{Mf} & $0$ \\
$F_2^0$ & $-$ & $0$ or $2$ & In Eq.~\ref{Mf} & $0$ \\
$F_3^0$ & $-$ & $0$ or $2$ & In Eq.~\ref{Mf} & $0$ \\
$S^\pm$ & $+$ & $8$ & $m_s^2+\frac{1}{2}y^2\phi^2$ & Ignored in resummation\\
$S^0$ & $+$ & $0$ or $4$ & $m_s^2+\frac{1}{2}y^2\phi^2$ & Ignored in resummation
\end{tabular}
\caption{Sign of contribution, degree of freedom, mass square and thermal mass square for each particle species in our model.}\label{V1particlelist}
\end{table}
The terms of $\bar{V}_1$ quartic in $\phi$ give corrections to the Higgs quartic coupling $\lambda$, which is equivalent to a direct RGE beta function calculation. It is clear from the supersymmetric point of view, that above the supersymmetric thresholds the fermionic negative contribution will just get canceled with the bosonic positive contribution. To study the threshold effect in more detail, we consider the box loop explicitly. The Higgs-charged fermion interaction term can be expanded in the $\psi,\chi$ basis
\begin{equation}\label{HiggsMix}
\frac{1}{\sqrt{2}}y\psi^{\mp(0)}\phi\chi^{\pm(0)}+\frac{1}{\sqrt{2}}y\chi^{\mp(0)}\phi\psi^{\pm(0)}=-\frac{1}{\sqrt{2}}y\sin2\theta F_1^{\mp(0)}\phi F_1^{\pm(0)}+\mbox{interaction with } F_2^{\mp(0)}.
\end{equation}
Summing over all the combinatorial of $F_1^\pm,F_2^\pm$ among the four propagators in the loop we get the total quartic coupling beta function coefficient proportional to 2, while the four light $F_1^\pm$ propagator only gives a contribution proportional to $\sin^42\theta$. With the beta coefficient step function approximation, between the scale of $M_{F_1^\pm}$ and $M_{F_2^\pm}$ the quartic beta function is only $\frac{\sin^42\theta}{2}$ the full strength. The situation is the same in the neutral fermion sector. So in the bottom up running the quartic coupling is corrected by
\begin{equation}\label{quarticRGE}
\Delta\lambda=-\frac{n_\pm}{64\pi^2}y^4\left(\frac{\sin^42\theta}{2}\ln\frac{M_S^2}{M_{F_1^\pm}^2}+\frac{2-\sin^42\theta}{2}\ln\frac{M_S^2}{M_{F_2^\pm}^2}\right)
-\frac{n_0}{64\pi^2}y^4\left(\frac{\sin^42\theta}{2}\ln\frac{M_S^2}{M_{F_2^0}^2}+\frac{2-\sin^42\theta}{2}\ln\frac{M_S^2}{M_{F_3^0}^2}\right).
\end{equation}
Requiring the fractional correction $\Delta\lambda<\lambda$ so that the quartic coupling remains positive, we can get a constraint on the boson physical mass $M_S$ and therefore on soft mass $m_s$. Further with $m_s$ we can determine the bosonic fractional diphoton amplitude change $\Delta{\cal A}_B$. \bigskip

\noindent\textbf{The finite temperature potential.} Taking finite temperature effect into consideration, we have new thermal corrections to the zero temperature potential, which are
\begin{eqnarray}\label{V1T}
V_{1T,i}&=&\pm n_i\int\frac{d^3p}{(2\pi)^3}T\ln\Big(1\mp e^{-\frac{\sqrt{\vec{p}^2+m^2_i}}{T}}\Big)
=\pm\frac{n_iT^4}{2\pi^2}J_{B/F}\Big(\frac{m_i}{T}\Big)\\
&=&\frac{n_iT^4}{2\pi^2}\times\left\{\begin{array}{ll}
{\displaystyle -\frac{\pi^4}{45}+\frac{\pi^2}{12}\frac{m^2_i}{T^2}-\frac{\pi}{6}\Big(\frac{m^2_i}{T^2}\Big)^{\frac{3}{2}}+\mathcal{O}\Big(\frac{m^4_i}{T^4}\Big)}&\mbox{Boson}\medskip\\
{\displaystyle -\frac{7\pi^4}{360}+\frac{\pi^2}{24}\frac{m^2_i}{T^2}+\mathcal{O}\Big(\frac{m^4_i}{T^4}\Big)}&\mbox{Fermion}
\end{array}\right.\nonumber
\end{eqnarray}
Still ``$+$'' is for boson and ``$-$'' is for fermion, and $T$ is the temperature. The second line assumes high temperature limit $T\gg m_i$, where the first term is the radiative pressure, in which fermions are corrected by a statistical factor of $\frac{7}{8}$; the second term is the thermal correction to mass square, and the third term in the bosonic potential is the cubic term which directly contributes to the strength of the EW phase transition.

Usually (for all the SM particles) $m^2\propto\phi^2$, so the $\frac{n_iT^4}{2\pi^2}\times\frac{\pi^2}{12}\frac{m^2}{T^2}$ ($\frac{\pi^2}{24}\frac{m^2}{T^2}$) term is also quadratic in $\phi$, and adds to the tree level negative Higgs mass term $-\frac{1}{4}m_h^2\phi^2$ a positive contribution proportional to $T^2\phi^2$ at high temperature, making the quadratic term positive and eliminating the spontaneous EW symmetry breaking. Taking the potential to be
\begin{equation}\label{thermalMassAndCubic}
V=D(T^2-T_0^2)\phi^2-ET\phi^3+\frac{1}{4}\lambda\phi^4,
\end{equation}
with key correction from thermal effect included, the cubic term contribution to the EW phase transition can be seen easily
\begin{equation}\label{SMStrength}
\frac{\langle\phi(T_c)\rangle}{T_c}=\frac{2E}{\lambda}.
\end{equation}
Here $T_c$ is the critical temperature where the potential at the newly developed local minimum $\phi\neq0$ is equal to the potential at $\phi=0$. The nucleation is usually right after the universe goes below the critical temperature, so we take $T_c\simeq T_n$ and this is a good approximate to the phase transition strength. In the SM $E=\frac{1}{6\pi v^3}(2m_W^3+m_Z^3)=6.4\times10^{-3}$ \cite{Dine:1992vs}, so the phase transition strength for $m_h\simeq125$~GeV is $\frac{2E}{\lambda}=0.1$ and therefore too weak for a successful EW baryogenesis\footnote{Numerically we have $T_n\simeq170$~GeV and $\langle\phi(T_n)\rangle\simeq20$~GeV, $\frac{\langle\phi(T_n)\rangle}{T_n}=0.12$.}.

With the expression of Eq.~(\ref{V1T}) we can see the most straightforward way to strengthen the phase transition is to add new strongly interacting bosons, which is exactly the new bosons in our model. A naive estimation shows if we simply ignore the soft term $m_s$ and plug in $m_i^2=\frac{1}{2}y^2\phi^2$, then pick out term cubic in $\phi$ in the expansion for bosons as part of $E$ term, for a strongly first order phase transition the new boson only need to contribute $n_iy^3\simeq6$, which implies $y\simeq1$. However, this estimation is wrong for two reasons: we have omitted both the thermal mass correction and the $m_s$ term, which make the high temperature expansion invalid. At finite temperature the boson (not the fermion) self energy loop will induce a correction proportional to $T^2$ to the mass square, which can be directly calculated through thermal Feynman diagram. A more efficient way to figure it out is through the above bosonic high temperature expansion Eq.~(\ref{V1T}), by comparing the quadratic term in the expansion to the tree level mass term. The SM particle thermal masses are well known. As for the new bosons, since we already have $m_s$ of order hundred of~GeVs even at Higgs VEV $\phi=0$ playing the same role as the thermal mass, and the thermal mass itself of the order $y^2T^2$, we expect its effect to be subdominant.% and can be ignored. On the other hand, to account for $m_s$ we will need a numerical calculation.
\bigskip

\begin{figure}
\centering
\includegraphics[height=2.5in]{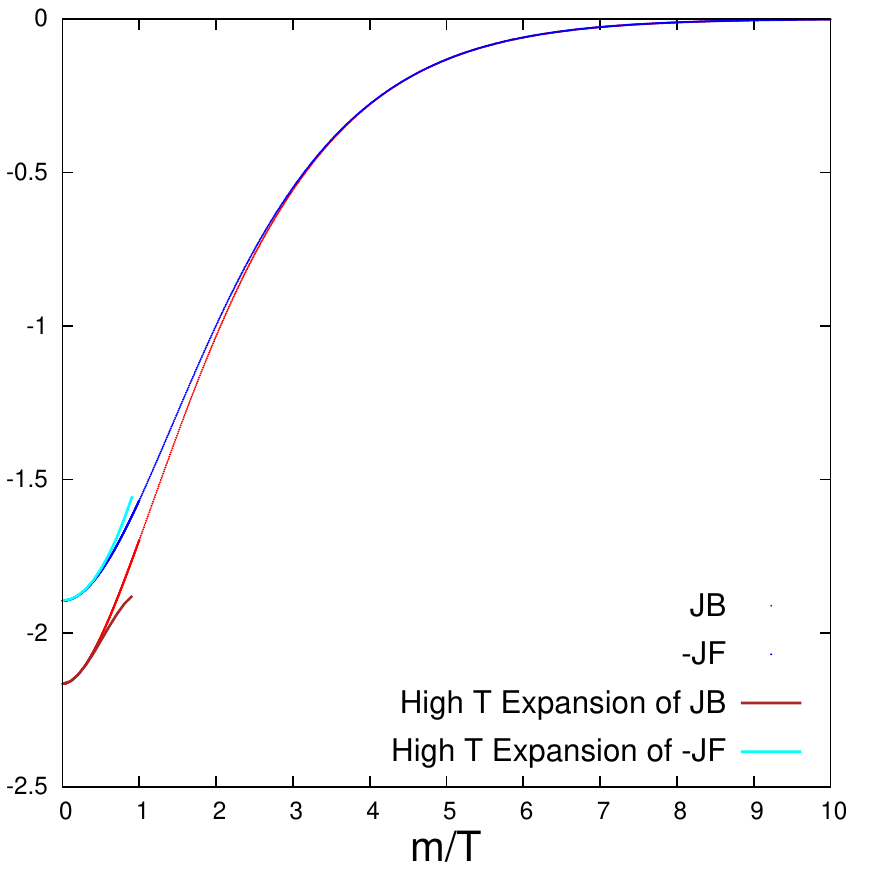}
\caption{The complete thermal one loop potential contribution $J_B$ (red curve) and $J_F$ (blue curve) and the comparison with their high temperature expansions (brown and cyan).}
\label{JBJF}
\end{figure}

\begin{figure}
\centering
\includegraphics[height=2.5in]{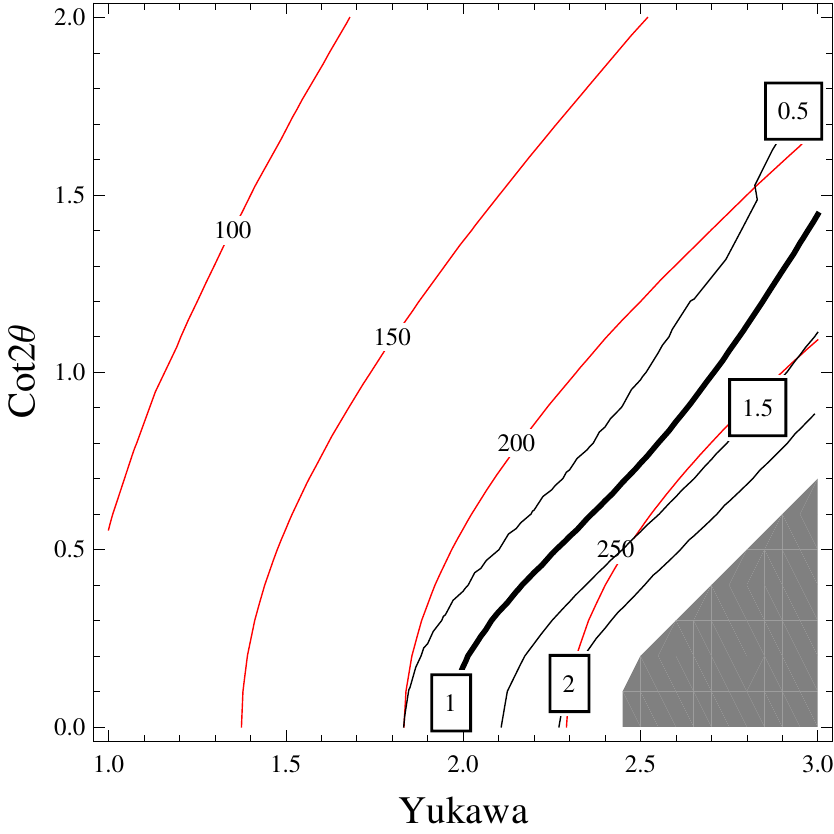}\qquad
\includegraphics[height=2.5in]{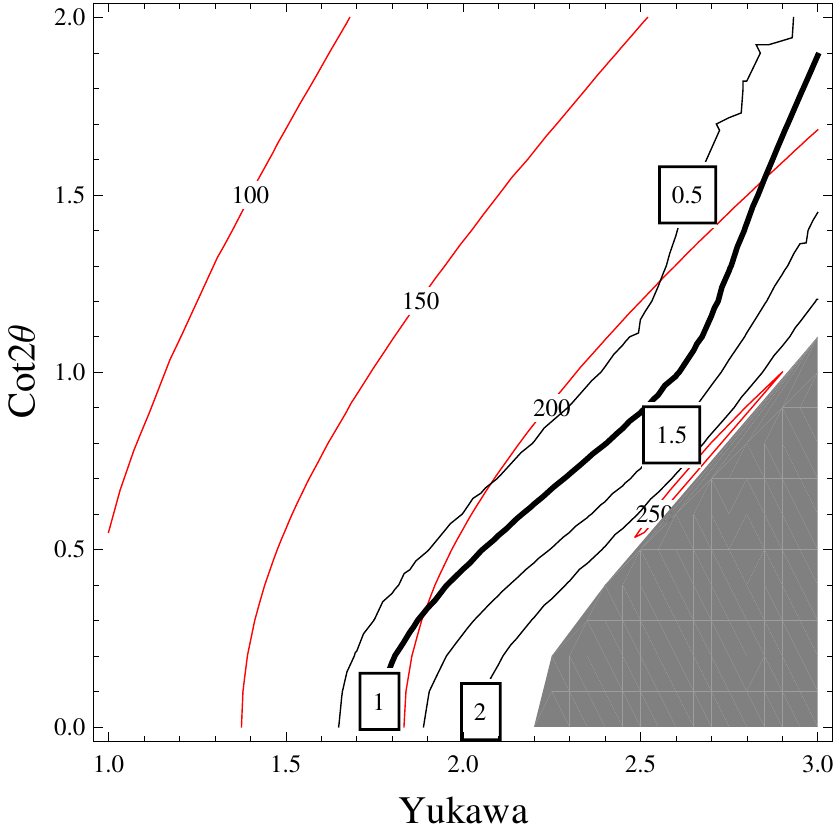}
\caption{The leptonic like $n_\pm=8, n_0=0$ (left) and wino-higgsino like $n_\pm=8, n_0=4$ (right) baryogenesis and lightest fermion mass contours. Here we choose $\Delta\mathcal{A}_F=0.25$ and $\Delta\lambda=0.5$. Black curves are EW phase transition strength $\frac{\langle\phi(T_c)\rangle}{T_c}$, and red curves are the lightest charged fermion mass. The gray shaded region are inconsistent with the EWSB for $T_c<0$.}
\label{da25}
\end{figure}

\begin{figure}
\centering
\includegraphics[height=2.5in]{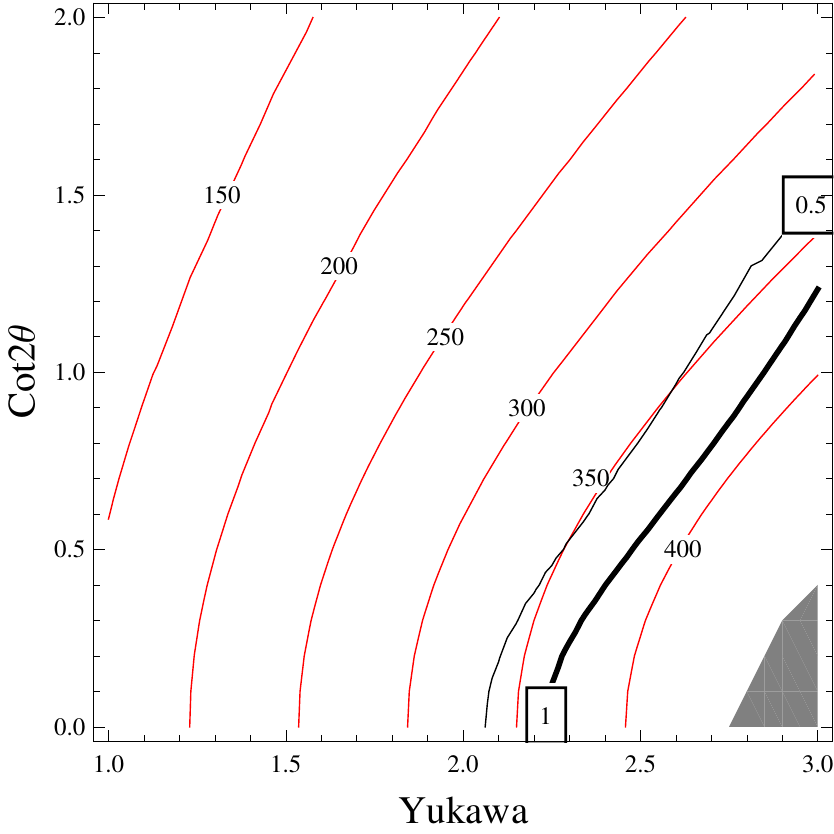}\qquad
\includegraphics[height=2.5in]{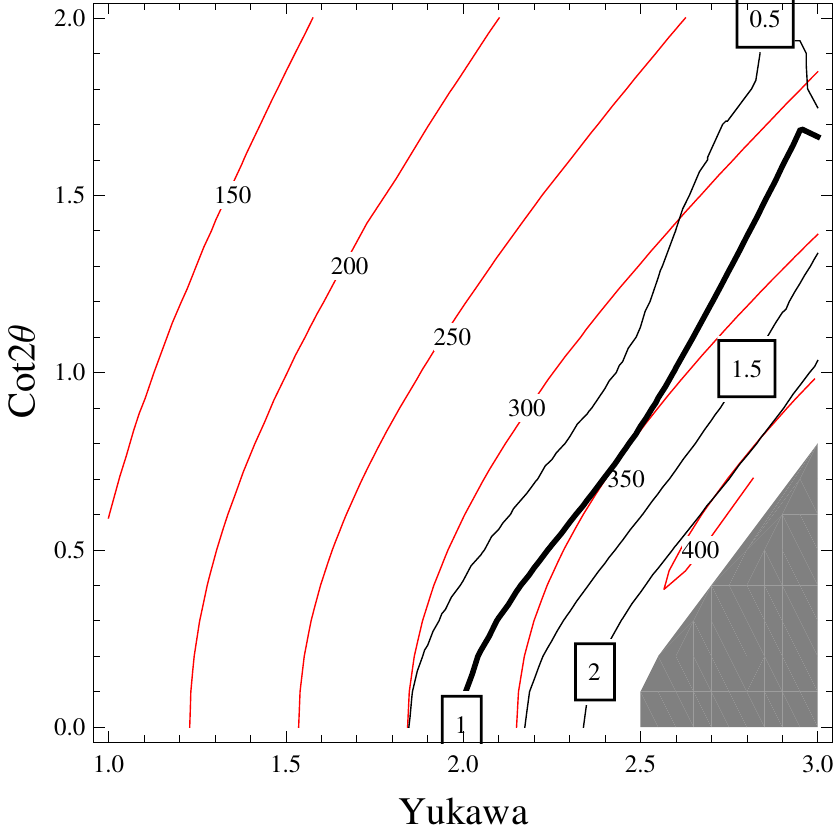}
\caption{Same as before, but for $\Delta\mathcal{A}_F=0.15$.}
\label{da15}
\end{figure}

\noindent\textbf{Results and Discussion.} We use the public code \texttt{CosmoTransition} \cite{Wainwright:2011kj} for a numerical evaluation of the phase transition, in which the finite temperature potential contribution $J_{B/F}$ are fully calculated as shown in Fig.~\ref{JBJF}, rather than relying on the hight temperature expansion. With fixed $n_\pm$, $n_0$ and desired $\Delta\mathcal{A}_F$ and $\Delta\lambda$, the soft masses $m_\psi$, $m_\chi$ and $m_s$ are functions of the fermion mixing angle $\theta$ and the effective Yukawa coupling $y$, as already shown in Eq.~(\ref{Mf}) and (\ref{quarticRGE}). The symmetry $\cot2\theta\leftrightarrow-\cot2\theta$ for the mass terms enable us to choose $\cot2\theta>0$. Our numerical results are shown in Fig.~\ref{da25} and \ref{da15}, separately for fermionic diphoton amplitude enhancement $\Delta\mathcal{A}_F=0.25$ and $\Delta\mathcal{A}_F=0.15$, with quartic coupling running both of $\Delta\lambda=-\frac{1}{2}\lambda_0=-0.0645$.


In the large $y$ and large mixing region we see a strongly first order phase transition. The effect of the new sector can be separated into bosonic and fermionic ones. In that region all the soft masses are a few hundreds~GeV. Given the largeness of the effective Yukawa coupling, significant part of the zero temperature mass comes from the zero temperature Higgs VEV. For example, in the leptonic case with $\Delta\mathcal{A}_F=0.25$, for $y=2.0$ and $\cot2\theta=0$ the scalar soft mass is $m_s=391$~GeV while the scalar mass $M_S=\sqrt{m_s^2+\frac{1}{2}y^2v^2}=524$~GeV, and the critical temperature is $T_c=133$~GeV. The new physics contribution to EW baryogenesis is mainly through the bosonic degree of freedom. The effect can be directly seen as a jump in the finite temperature potential. Although the zero temperature Boltzmann suppression of the soft mass of $e^{-\frac{m_s}{T_c}}$ makes the value at $\phi=0$ on the $J_B$ curve of $x=3$ much lower than that of $x=0$ in magnitude, the jump from $x=3$ to $x=4$ at $\phi\neq0$ is still significant. Out of the general argument in \cite{Chung:2012vg}, here we manage to find a region in which the Boltzmann suppression does not completely spoil the contribution of the new coupling. This novel effect with large soft mass at $\phi=0$ is different from the light stop scenario \cite{Carena:1996wj} or its variant \cite{Huang:2012wn}, which is still around $\frac{m}{T}\sim0$ and described by the high temperature expansion.

On the other hand, the fermionic degree of freedom contributes negatively to phase transition strength. This is because in order to provide the desired diphoton amplitude shift, our fermionic mass matrix has an inverted dependence of the light fermion mass on the Higgs VEV. When a nonzero Higgs VEV is turned on, the light fermion which dominates both the diphoton loop amplitude and the baryogenesis contribution becomes even lighter, so the jump on the $J_F$ curve is effectively backward, partially compensating the contribution of bosons. This contribution is roughly half of the bosonic contribution, because the light fermion degree of freedom is halved compared to the boson. So in our model with fermions alone we will never get successful EW baryogenesis. This follows the general argument in \cite{Chung:2012vg} that the diphoton excess itself requires a Higgs portal coupling opposite to the one required by EW baryogenesis, but we have a way out by the more significant bosonic contribution.

To achieve larger EW phase transition strength one can increase the jump of the Higgs VEV or lower the transition temperature. In fact, both happen in our model. The bound for the Higgs VEV jump is constrained by the zero temperature Higgs VEV, and no jump larger than $v=246.2$~GeV is available. On the other hand, as the effective Yukawa goes larger and the mixing goes larger, the critical temperature for EW baryogenesis gets smaller due to the correction from the  $\frac{n_i}{64\pi^2}2m_i^2(v)m_i^2(\phi)$ renormalized potential term of the bosonic degree of freedom. It happens in our model all the way until the physical requirement $T_c>0$ is spoiled, which is shown as the gray shaded regions. Too low a critical temperature may create problems for the tunneling to occur, however, we can simply be satisfied with a slightly larger than one phase transition value. For simplicity we choose to plot the results for $T_c>100$ GeV.

\begin{figure}
\centering
\includegraphics[height=2.5in]{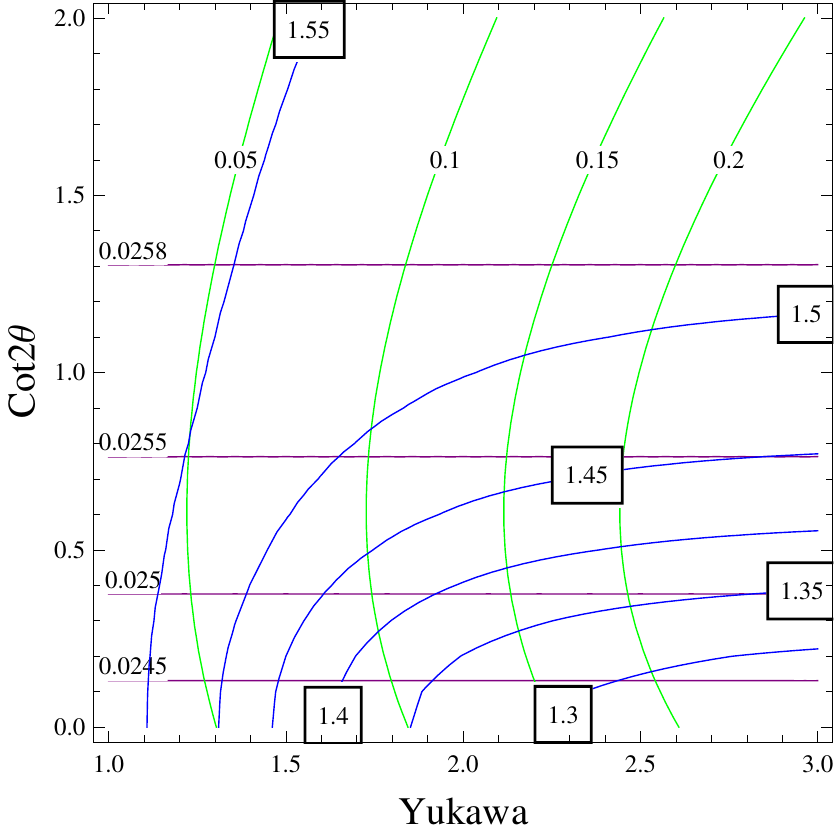}\qquad
\includegraphics[height=2.5in]{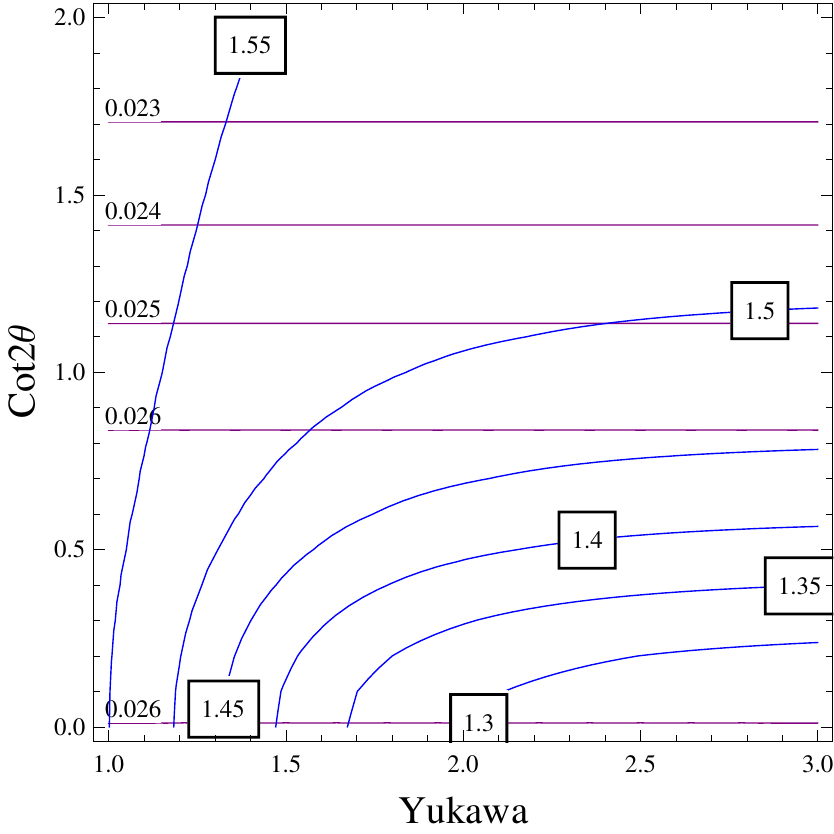}
\caption{The leptonic like $n_\pm=8, n_0=0$ (left) and wino-higgsino like $n_\pm=8, n_0=4$ (right) Higgs diphoton decay branching ratio and EW precision observable contours. Here we choose $\Delta\mathcal{A}_F=0.25$ and $\Delta\lambda=0.5$. Blue curves are diphoton signal strength $\mu_{\gamma\gamma}$, green curves are Peskin - Takeuchi $T$ parameter and purple curves are $S$ parameter. In the wino-higgsino case the $T$ parameter vanishes.}
\label{da252}
\end{figure}

\begin{figure}
\centering
\includegraphics[height=2.5in]{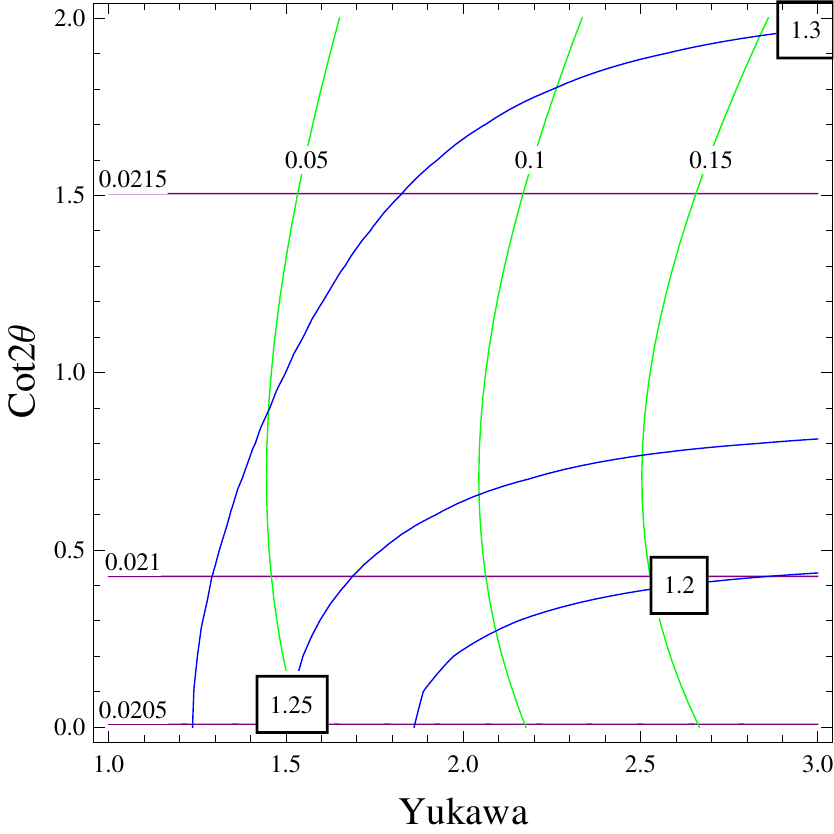}\qquad
\includegraphics[height=2.5in]{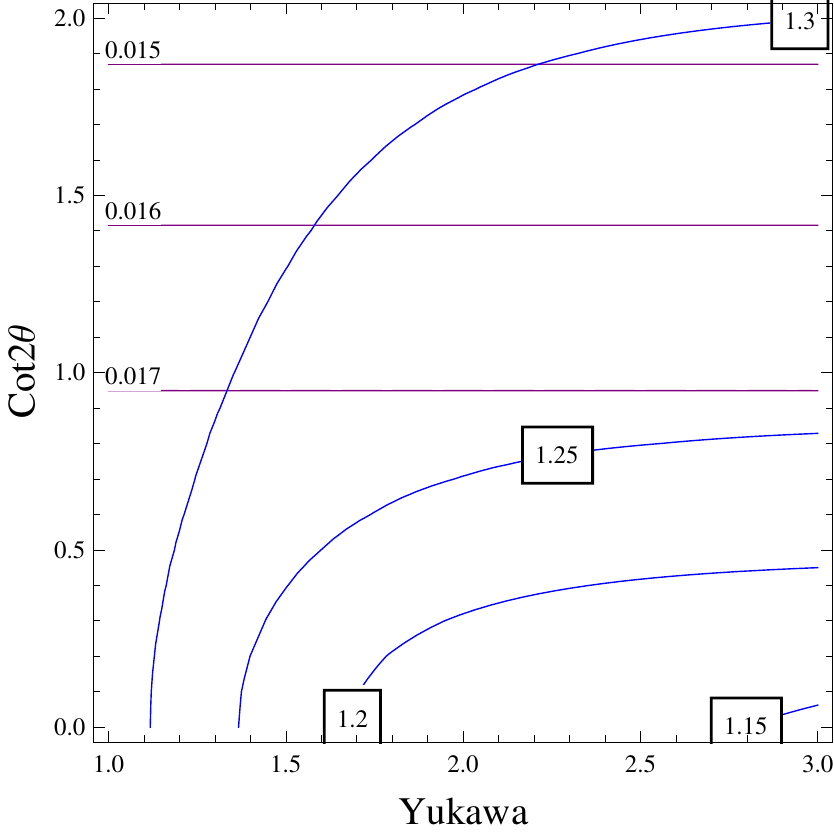}
\caption{Same as before, but for $\Delta\mathcal{A}_F=0.15$.}
\label{da152}
\end{figure}

As for the diphoton rate, the bosonic component contribution compensates the fermionic amplitude contribution. The compensation is more efficient in the low $m_s$ and successful EW baryogeneisi region, which are shown in Fig. \ref{da252} and \ref{da152}. For $\Delta\mathcal{A}_F=0.25$ usually we get $\mu_{\gamma\gamma}$ of $1.3$ to $1.4$, and for $\Delta\mathcal{A}_F=0.15$ we get $\mu_{\gamma\gamma}$ about $1.2$. Since $\Delta\mathcal{A}_F$ is free parameter in the model, $\mu_{\gamma\gamma}$ can be tuned freely.

The lightest new particle in our model is always the light charged (or neutral) fermion, the mass of which is two hundred to four hundred GeV in the interesting region, as shown in Fig.~\ref{da25} and \ref{da15}. Some discussion of the collider signature can be found in \cite{ArkaniHamed:2012kq}. In our model because we are interested in a high effective Yukawa coupling region, the vector like masses also get higher, allowing the model to escape from the current collider mass bound constraints. Detailed collider search analysis will be reserved for a separated work.

In the same figures we have also calculated the Peskin - Takeuchi $T$ and $S$ parameters. Since the bosonic components are always degenerated in our model, they do not contribute to the oblique correction. For the wino-higgsino model we take the hypercharges of the two higgsinos to be different, namely the MSSM like one. Then it preserves a custodial symmetry, so the $T$ parameter contribution vanishes. We can see in the interesting parameter region, the $T$ and $S$ parameters are consistent with the experimental constraint \cite{pdg}.


% In this paper we are focusing on getting a strongly first order phase transition. The CP violation phases are discussed \cite{Shu:2013uua}.

\noindent\textbf{Acknowledgement.} The author is grateful to discussion with Carlos~E.~M.~Wagner and Jing Shu.

\end{document}